\documentstyle[11pt]{article}
\begin{document}

\title{Generalized Ohm's Law}
\author{Yuli V. Nazarov}
\centerline{Faculteit der Technische Natuurkunde,
Technische Universiteit Delft,}
\centerline{ 2628 CJ Delft, the Netherlands}
 \date{22.09.94}

In 1827 George Simon Ohm formulated his law thus pioneering
studies of electric conductivity.

In 1970 Rolf Landauer \cite{Landauer} has published a contribution
establishing a modern {\it scattering} approach to the conductivity.
In the framework of this approach, a conductor
is treated as a single complicated scaterrer.
It is feasible for {\it coherent}
conductors,
where resistive region so small that
 no inelastic process
takes place when a carrier traverses the region.

In the last quarter of the 20th century
such discoveries as weak localization, universal conductance
fluctuations and conductance quantization in ballistic regime
have expanded by far our knowledge of quantum transport in coherent conductors.
The scale of the abovementioned effects comprises several conductance
quanta $G_Q=e^2/(2 \pi \hbar)$.
This is why those are marginal for classical
conductors having conductance $\gg G_Q$.

In 1992 Beenakker and B\"{u}ttiker   \cite{onethird} discovered that
the excess noise power in coherent diffusive {\it classical} conductor
comprises exactly one third of its {\it classical} estimation.
In my opinion, the importance and impact of this work has not  yet been
fully appreciated. This is a new finding in classical conductivity,
in the area which seemed to be exhaustively explored shortly after Ohm's
discovery. This work explicitly shows that to complete our knowledge
of classical conductivity, we  have to know a scattering
characteristic of a conductor: its  {\it transmission
distribution}.

The purpose of this article is to give a set of rules
to compute this transmission distribution.
Those are based on solid microscopic background. They are no
more complicated than Ohm's law and possess in fact
 the similar structure. We call it {\it generalized Ohm's law}.

  This work is intended to be comprehensive and is structured
  as follows. In introductory section I remind the reader the basics of
  Ohm's approach and scattering approach to conductivity. The second section
presents
  a microscopic derivation of the generalized Ohm's law and shall be skipped
  by everyone who is not interested in such details. Diffusive
  conductors are discussed in the third section.
  The important fourth section contains a concise formulation of the
generalized
  Ohm's law in terms of a circuit theory and shows how to deal with tunnel
junctions and quantum point contacts.
   Examples and results are to be found in the last section.

\section{Introduction}
\subsection{Ohm's law}
In its global form, Ohm's law relates the total current in a conductor
to the potential drop over it,
\begin{equation}
I=GV,
\end{equation}
$G$ being conductivity. It is valid for any conductor at small voltages and is
in fact trivial.
The local form of Ohm's law contains more
information. It relates local current density and
 local voltage gradient,
\begin{equation}
j^{\alpha}(x)=-\sigma(x)
\frac{\partial}{\partial x^{\alpha}} V(x).
\end{equation}
The conservation of current implies
that $ {\rm div} {\bf j}=0$. Therefore voltage distribution
over the conductor obeys
\begin{equation}
\frac{\partial}{\partial x^{\alpha}}\sigma(x)
\frac{\partial}{\partial x^{\alpha}} V(x)=0.
\label{Laplace}
\end{equation}
 It must also satisfy boundary conditions
\begin{eqnarray}
V(x)=0,  {x \rightarrow -\infty} & V(x)=V,
{x \rightarrow \infty} \\
N^\alpha (x) j^\alpha (x)=0, & x \subset {\rm boundary},
\label{boundary}
\end{eqnarray}
$N^\alpha(x)$ being vector normal to the boundary in a given
point $x$. The latter condition can be extended to describe
interfaces in the conductor if any:
\begin{equation}
N^\alpha(x) j^\alpha(x)= g(x) (V_1(x)-V_2(x)),
\label{interface}
\end{equation}
here unit area conductivity of the interface $g(x)$
relates the current density and voltage drop at the
interface.

Eq. \ref{Laplace} with boundary condition determines
the total conductance: the total current through
an arbitrary cross-section of the conductor equals
$G V$.

The calculation of conductance is a practical problem
of elecroengeering.
We would have used candles to light if we had
to solve differential equations for such calculation.
Indeed, there is a finite element approximation of Eqs. \ref{Laplace},
\ref{boundary}, \ref{interface} known as {\it circuit theory}.
The conductor is separated in finite elements connected
in nodes. Conservation of the current in nodes with current-
voltage relations for each element set the voltage in each
node and finally determine the total conductance.

\subsection{Transmission matrix}

Alternatively,  electronic conductance can be
treated as a complicated elastic quantum scattering problem.
The conventional way to start is to regard terminals
of the conductor as  semi-infinite perfectly conducting leads of finite width.
Due to quantization of transverse momentum,
there is a finite set of propagating channels in the leads
at given energy. Transmission matrix $t_{nm}$ is defined
in the basis of incoming channels  in one of the wires
and outgoing channels  in another wire. If the incident
beam is concentrated in the channel $n$, the wave function
in another wire takes the form $\sum_m t_{nm} \vert m >$.

Landauer formula expresses the conductance in terms
of eigenvalues of the  transmission matrix square,
${\bf T}={\bf t}{\bf t^{\dagger}}$,
\begin{equation}
G= G_Q {\rm Tr} {\bf T}=
G_Q \sum_n T_n.
\label{Landauer}
\end{equation}
Transmissions $T_n$ are distributed from 0 to 1.

Specific values of $T_n$'s are sensitive to the configuration
of impurities in the sample. To characterize a big
classical conductor, we introduce transmission distribution
averaging over all possible impurity configurations,
\begin{equation}
\rho(T)= \sum_n <\delta(T-T_n)>.
\end{equation}

In a disordered conductor all channels seem alike, and one
might assume comparable transmissions $T_n \simeq l/L$ for
all channels, $l$ and $L$ being mean free path and the sample
size respectively.
The latter estimation would follow from the comparison with Ohm's law.

As it was first shown by Oleg Dorokhov \cite{Dorokhov}, this
is completely wrong. In fact, most of the channels are "closed" possessing
exponentially small transmission. The transport is due to a few
"open" channels \cite{openchannels} having transmissions of the order of unity,
and the distribution of those,
\begin{equation}
\rho(T) \equiv <\sum_n \delta(T-T_n)> =
\frac{G}{2 G_Q} \frac{1}{T \sqrt{1-T}} \equiv \rho_0(T),
\label{result}
\end{equation}
is universal. Such an unusual fact concerning matrix eigenvalues
should be comprehended in the framework
of a certain random matrix theory, and
the relation (\ref{result}) can be proven by random matrix
methods \cite{Mello}. Let us stress that both methods essentially exploit
an assumption of a uniform
quasi-one-dimensional conductor,that is, are
formally valid only if the length of the disordered region
by far exceeds its width.

It has been proven in \cite{Universality} that Eq. \ref{result}
holds for a diffusive conductor of an arbitrary shape.
It was indicated there that this result does not hold
if the scattering is not diffusive in a part of the sample.
It is the case when there are extended defects in the
conductor such as grain boundaries, tunnel barriers, cracks.
It is the case when the resistivity of the sample at
least partially determined by ballistic transmission, that is,
there are (quantum) point contacts in the circuit.

\subsection{THe use of transmissions}

At first sight, the transmission distribution does not seem
very practical. We know the conductance of the conductor
\begin{equation}
G= G_Q \int_0^1 dT T \rho(T)
\end{equation}
anyhow, why should we know the detailed shape of $\rho(T)$?

Recently it has been recognized that many {\it other}
transport properties of the conductor can be expressed
in terms of transmission distribution. We list them
below.

B\"{u}ttiker \cite{ButtikerNoise} and Lesovik \cite{Lesovik}
have found that the magnitude of a shot noise in a
conductor may be expressed in terms of transmissions:
\begin{equation}
P= 2 G_Q \vert eV\vert \sum_n T_n (1-T_n).
\end{equation}
Classical Poisson noise $P_{Poisson}= 2eI$ would correspond to the first term
in brackets. It is convenient to characterize the shot noise
power by the suppression factor
\begin{equation}
P/P_{Poisson}= 1- \sum_n T_n^2/\sum_n T_n.
\label{noise}
\end{equation}

Lesovik and Levitov \cite{Binomial} managed to provide a
complete description of the temporal current fluctuations.
They found that
the number of electrons having passed the conductor
in channel $n$
obeys binomial statistics, probability $P_K$ for
$K$ electrons to pass for the time $t_0$ being
\begin{equation}
P_K= T_n^K (1-T_n)^{N-K} C^K_N.
\label{noisestatistics}
\end{equation}
Here $N\equiv eV t_0$, $C^K_N$ being
binomial coefficients.

Beenakker has found the scattering approach to be useful
in superconductivity.
Conductance of the normal conductor in series with
a superconductor is due to Andreev reflection. Such
Andreev conductivity is given by \cite{CarloAndreev}
\begin{equation}
G_A=2 G_Q \sum_n \frac{T_n^2}{(2-T_n)^2}
\label{Andreev}
\end{equation}

 If a normal conductor is placed between two
superconductors, non-dissipative Josephson
current may flow through the system. It depends
on  phase difference $\phi_{sup}$ between superconductors
and is given by \cite{CarloJosephson}
\begin{equation}
I_J(\phi_{sup}) = \frac{\pi \Delta \sin \phi_{sup}}{2 e} G_Q
\sum_n \frac{T_n}{\sqrt{1- T_n \sin^2(\phi_{sup}/2)}}
\label{supercurrent}
\end{equation}
$\Delta$ being the energy gap in the superconductors.
These two formulas are approximate being only suitable
 for small structures at low temperatures.
The limits of their validity are discussed in
\cite{CarloAndreev}, \cite{CarloJosephson}.

Averaging Eqs. \ref{noise}, \ref{Andreev}, \ref{supercurrent}
with a proper transmission distribution, we are able to
find the corresponding transport properties of the
conductor.

Very recently Nieuwenhuizen and van Rossum \cite{Nieuw}
have found an important relation between the transmission
distribution and intensity distribution of waves
transmitted through a multiple scattering medium.
The latter can be directly measured by optical
means.

Since the importance and physical relevance
of the transmission distribution is established,
it is worthwhile to calculate it from the
standard microscopic model.

\section{ Microscopic method for transmission matrix}

\subsection{Green's functions and transmission}
 Green's function technique for disordered conductors
have been elaborated in detail (see, e.g. \cite{Review})
and we will use this one. Therefore,
as a first step we have to establish the relation between Green functions
and the  transmission matrix amplitudes. Let us consider first a disordered
region with two perfect leads attached. Transverse motion in the leads is
quantized giving rise to discrete modes. The traversal motion
along the coordinate $z$ is not quantized.
Electron Green function $G^{A(R)}_{nm}(\epsilon;z,z')$, $n,m$ being transverse
mode indexes,
describes an evolution of a wave packet at energy close to Fermi level
that stars at point $z (z')$. Let us take two cross-sections of the
leads, one far to the left and another one to the right. At the cross-sections,
the electron wave functions are given by asymptotes of scattered waves.
This allows to write for transmission from  left to right:
\begin{equation}
t_{mn}=i\sqrt{v_n v_m} G^{A}_{mn}(z,z'),
\end{equation}
where $z,z'$ belong to the left and right cross-sections respectively.
The conjugated matrix is expressed through $G_R$.

We are ready to find expressions for ${\rm Tr}({\bf t^{\dagger} t})^n$.
We replace summation over the quantized modes  by integration over
the transverse coordinates. Green function should be written in the
coordinate representation, thus given by ($x$ is three-dimensional coordinate)
\begin{equation}
(\epsilon \pm i \delta -\epsilon(\hat{p_1})  -U(x)) G^{R,A}(x_1,x_2)=
\delta(x_1-x_2),
\label{zerogreen}
\end{equation}
$U(x)$ being random impurity potential.

For the lowest order trace this yields
\begin{equation}
{\rm Tr}({\bf t^{\dagger} t}) = \int d^3 x_1 d^3 x_2 d^3 x_3 d^3 x_4
v_1(x_1,x_2) G^{A}(x_2,x_3) v_2(x_3,x_4) G^{R}(x_4,x_1).
\label{Kubo}
\end{equation}
Here $v_{1(2)}(x,x')$ stands for the operator of the current through
left (right) cross-section.
Eq.(\ref{Kubo}) contains a correlator of two current operators
and it is easy to see that this is equivalent to Kubo formula
for conductivity. All traces of this kind possess the same operator
structure, that we write symbolically as
\begin{equation}
{\rm Tr}({\bf t^{\dagger} t})^n={\rm Tr}_{x} (v_1 G^{A} v_2 G^{R})^n.
\label{traces}
\end{equation}
It seems like (\ref{traces}) shall depend on a choice of cross-sections,
but it is not true. This can be checked with (\ref{zerogreen}) and follows from
conservation law for the current. Moreover, this does not depend on the shape
of a cross-section and whether it is in a lead or in a disordered region.
In this way, one can relax the requirement of perfect leads
(which is the weakest point of Landauer formalism).

\subsection{Multicomponent Green's function}
In this way we can compute the trace of every power of {\it \bf T}
but our goal is to find the transmission distribution.
The idea is to relate a sum of (\ref{traces}) taken with certain weights,
first, to the transmission distribution, second, to a trace
of a certain multicomponent Green function in the field of a fictitious
potential. This potential couples advanced and retarded functions,
 therefore
resulting Green function must be two-by-two matrix labeled by  indexes
$i,j=A,R$. We introduce such Green function by
the following equation: (hats denote two-by-two matrices)
\begin{eqnarray}
\hat G(x_1,x_2)=\hat G^{(0)}(x_1,x_2)+
\int  d^3 x_3 d^3 x_4\hat G^{(0)}(x_1,x_3) \nonumber \\
(\zeta_1 v_1(x_3,x_4)  \hat\tau^{\dagger}+
\zeta_2 v_2(x_3,x_4)  \hat\tau) \hat G(x_4,x_2),
\label{Green}
\end{eqnarray}
$\zeta_{1,2}$ characterizing the fictitious potential
at the crosssections.
Here the following matrices have been used:
\begin{equation}
\hat G=\left(\begin{array}{c} G^A \ \ 0 \\ 0 \ \ G^R \end{array}\right);
\hat \tau=\left(\begin{array}{c} 0 \  1 \\ 0 \  0 \end{array}\right);
\end{equation}
$G^{A,R}$ satisfying (\ref{zerogreen}).
By standard Green's function methods we can derive
the key result of the microscopic approach:
\begin{equation}
\int d^3 x_1 d^3 x_2 v_1(x_1,x_2) {\rm Tr} (\hat\tau^{\dagger} \hat G(x_1,x_2))
=
\zeta_2 \rm{Tr} (\frac{{\bf t^{\dagger} t}}
{1-\zeta_1\zeta_2 {\bf t^{\dagger} t}}) \equiv \zeta_2 F(\zeta_1\zeta_2)
\label{thefunction}
\end{equation}
It is easy to check that Taylor series of (\ref{thefunction}) in
$\zeta_1\zeta_2$ will give operators in
(\ref{traces}). We see that the function $F(x)$ may be
an immediate output of microscopic calculation.
The transmission distribution will be related to $F(x)$ at complex $x$.

\subsection{Semiclassical equations}
The wisdom of Green's function technique allows one to
 average easily the Green's  function in use
over configurations of scatterers. In semiclassical
approach we are in it can be expressed as
\begin{eqnarray}
<\hat G(x_1,x_2)> =\hat G^{(0)}(x_1,x_2)+
\int  d^3 x_3 d^3 x_4\hat G^{(0)}(x_1,x_3) \nonumber \\
(\zeta_1 v_1(x_3,x_4)  \hat\tau^{\dagger}+
\zeta_2 v_2(x_3,x_4)  \hat\tau +\hat\Sigma_{imp}(x_3,x_4) ) <\hat G(x_4,x_2)>,
\label{averaged}
\end{eqnarray}
The impurity selfenergy $\Sigma_{imp}$ may take
a simple form for white noise random potentilal
\cite{Review}
\begin{equation}
\hat\Sigma_{imp}(x,x') = \frac{1}{2 \pi\tau(x)} \hat G(x,x') \delta(x-x')
\end{equation}
$\tau(x)$ being (space-dependent) electron transport
time, $\nu$ being density of states near
Fermi level per one spin direction.

In the following we will keep one of the most
beaten tracks of condensed matter physics.
It starts from  Green's function relations
similar to (\ref{averaged}) that treats
scatterers in microscopic level.
It passes the stage of Boltzman master equation
where information about scatterers is presented
in form of transition probabilities.
It finally comes to diffusion equations where
all information about scattering is compacted
into diffusion coefficient $D(x)$.
It is a straight way for one component
Green's function, we shall do with a multicomponent
one. Fortunately, the multicomponent Green's functions
have been explored in connection with superconductivity,
and technique we use exhibits many similarities with
the one developed to study non-equilibrium properties of
superconductors \cite{Larkin}.
Refs. \cite{Larkin}, \cite{LarkinOhm} show in detail how
to deal with multicomponent Green's functions
at every turn of the track,
and I will skip  technicalities referring a reader
to these papers.

The derivation ends up with an effective diffusion
equation for the averaged Green function in coinciding points, $
\hat G(x,x) \equiv i\pi\nu \hat\Lambda(x)$.  The matrix $\hat\Lambda$ obeys
unitary
condition $\hat\Lambda^2=\hat 1$. The diffusion
coefficient can be expressed in terms of local conductivity
$\sigma(x)$, $D(x)=\sigma(x)/(2 e^2 \nu)$.
The resulting diffusion equation can be written
as a conservation law for  matrix current:
\begin{eqnarray}
\frac{\partial
j_{\alpha}(x)}{\partial x_\alpha}=0; &
j_{\alpha}(x)=\sigma(x) \hat{\Lambda}^{-1}
\frac{\partial
\hat{ \Lambda}}{\partial x_\alpha}.
\label{diffusion}
\end{eqnarray}
The fictitious potential may be incorporated into boundary conditions
at left and right infinity:
\begin{eqnarray}
\hat\Lambda(-\infty)=\hat\sigma_{z}; &
\hat \Lambda(\infty)=
\hat S \hat \Lambda(-\infty)\hat S^{-1} \\ \label{lambda_bound}
{\rm where \ \ } &
\hat S(\zeta_1,\zeta_2)=\exp(i\zeta_1\hat\tau^\dagger)\exp(i\zeta_2\hat\tau),
\end{eqnarray}
so that the solution of (\ref{diffusion}) explicitly does not depend
upon the choice of cross-sections. The same is true for the value of
interest(\ref{thefunction}) which can be expressed in terms of the total matrix
current
through an arbitrary cross-section of the conductor as follows:
\begin{equation}
\zeta_2 F(\zeta_1 \zeta_2) = i \pi  {\rm Tr} \hat \tau^{\dagger} \hat I/G_Q
\label{for_f}
\end{equation}
\[ \hat I \equiv \int_S \hat j_\alpha(x)N_\alpha(x),\]
$N_\alpha$ being normal vector to the cross-section surface.
The result does not depend on the cross-section since the current conserves.

At arbitrary $\zeta_1,\zeta_2$  Eq. (\ref{diffusion}) is in general
a complex nonlinear matrix equation.
 Fortunately, the solution we are
searching for can be found easily.
 Let us parameterize $\hat\Lambda$
as follows:
\begin{equation}
\hat {\bar\Lambda} = \left(\begin{array}{c} \cos\theta \ \  \sin\theta/B\\
B \sin\theta  \ \ -\cos\theta \end{array}\right); \ \
B \equiv -i\sqrt{\frac{\zeta_2}{\zeta_1(1-\zeta_1\zeta_2)}}
\end{equation}
Under this parameterization, the diffusion equation (\ref{diffusion})
becomes linear,
\begin{equation}
\hat j_{\alpha}=\sigma(x) \frac{\partial \theta}{\partial x_\alpha}
\left(\begin{array}{c} 0 \ \ 1/B \\ B \ \ 0 \end{array}\right);
\frac{\partial}{\partial x_\alpha} \sigma(x)
\frac{\partial \theta}{\partial x_\alpha} =0.
\label{theta}
\end{equation}
It is convenient to choose $\zeta_1=\zeta_2=\sin(\phi/2)$
and introduce by definition the following quantities
\begin{eqnarray}
F(\phi) \equiv  \rm{Tr} (\frac{{\bf t^{\dagger} t}}
{1- \sin^2(\phi/2) {\bf t^{\dagger} t}}),  &
I(\phi) \equiv G_Q \sin \phi \rm{Tr} (\frac{{\bf t^{\dagger} t}}
{1- \sin^2(\phi/2) {\bf t^{\dagger} t}})
\label{definition}
\end{eqnarray}
to characterize transmission distribution.
The distribution of transmissions can be extracted from $F(\phi)$
or $I(\phi)$ in a complex
plane of $\phi$. Using the identity $\pi\delta(x)={\rm Im}((x-i0))$ we show
that
\begin{equation}
\rho(T)= \frac{\rho_0(T)}{\pi G}\ \  {\rm Re}(I(\pi + i\  {\rm arccosh}
(\frac{1}{\sqrt{T}}))).
\label{analytic}
\end{equation}

Boundary conditions (\ref{lambda_bound})
take then a simple form
\begin{equation}
\theta(-\infty)=0, \theta(\infty)=\phi
\label{theta_b}
\end{equation}
 Making
use of Eqs. \ref{for_f}, \ref{definition} we prove
that
\begin{equation}
I(\phi)= -\int_S\sigma(x) N^{\alpha} \frac{\partial \theta}{\partial
x_{\alpha}},
\label{tot-tot}
\end{equation}
the integral may be taken over an arbitrary
crosssection of the sample.

\section{Universality of diffusive conductors}
 In the previous section we developed a microscopic
approach to transmission matrix and found that
the function $I(\phi)$ that characterizes the
transmission distribution can be determined
from Eq.\ref{theta}, boundary condition (\ref{theta_b}), by
means of relation \ref{tot-tot}. I shall stress that
$I(\phi)$ is a function of transmission matrix
defined by (\ref{definition}). By no means it is
an electric current.

However let us recognize a striking similarity
between (\ref{theta}) and Laplace equation
(\ref{Laplace}) that express Ohm's law.
Now we can immediately relate $\theta(x)$ to
potential distribution in the conductor biased
by the potential difference $\phi$. Therefore,
$I(\phi)$ equals the total current through
the conductor biased by $\phi$ and
\begin{equation}
I(\phi) = G \phi
\end{equation}
 by definition of conductance. Applying (\ref{analytic})
we obtain
\[\rho(T)=\rho_0(T).\]
So  we have proved the validity of(\ref{result}) for an
arbitrary diffusive conductor. This is because the problem can be
related to the solution of the equation determining the voltage distribution
in the conductor.

\section{Circuit theory}

\subsection{Circuits of diffusive conductors}

We saw in the previous chapter that
the microscopic problem of finding the transmission
 distribution can be reduced to  a simple form, that
resembles much the procedure of finding the sample
conductance if the local conductivity is known.

Indeed, we have a "voltage drop" $\phi$
over the conductor and distribution of local
"potentials" $\theta(x)$. Local "current" density
is related to local gradient of $\theta(x)$.
Conservation of the "current" enables us to find
the "potential" distribution. If we know the latter
we can calculate the total "current" $I(\phi)$ through
the sample. The last step is specific: we shall determine
the transmission distribution from $I(\phi)$ by
making use of Eq. \ref{analytic}.

This prompts  a further simplification which
is well-known from electric conductivity theory.
To describe complicated, compound, artificially
designed systems we can use a kind of circuit theory.
We will consider only two terminal circuits.
Let us separate a conductor into finite elements
connected in nodes. Each node has a certain "potential".
The "current" through each
element is given by the "potential" difference
dropping on it. The  conservation of the
"current" in nodes enables us to find the
total "current" $I(\phi)$ and to solve the transmission
distribution for the network given. The rigorous
circuit theory rules will be formulated below.
Let us first understand why  we need those.

Indeed, if we assemble the circuit from the diffusive
conductors there is no problem whatsoever.
The transmission distribution is "universal"
depending on the total conductivity only.
In circuit language, this is a property of
a {\it linear} circuit($I(\Delta \theta)=G \Delta \theta$ for diffusive
elements): if "current" is
directly proportional to the voltage drop
for each element, the same is true for the
whole circuit.

The  situation is different if some of
the elements possess non-linear $I-\theta$
characteristic. What are these special
circuit elements? Do they occur in reality?

\subsection{Special circuit elements}

The answer is positive. If we consider a realistic
microstructure, very often it contains non-diffusive
elements. There may be tunnel junctions partitioning
off the conductor. The resistivity of the structure
may be dominated by ballistic transmissions,
like in (quantum) point contact arrays. There is
a great current interest in hybrid structures
of different kinds. Combining diffusive conductors,
tunnel junctions and ballistic constrictions,
one can obtain a variety of coherent devices.
The circuit theory we present can deal with
those.

We show below that the tunnel junctions and
ballistic constrictions possess non-linear
$I-\theta$ characteristic. To avoid confusion,
I shall stress that their $I-V$ curves are
supposed to be linear. To obtain $I-\theta$
characteristics, we start immediately from
the definition (\ref{definition}).

Let us consider first  a tunnel
junction. If the tunnel barrier is thick enough,
the transmissions are much smaller than unity.
{}From Eq. \ref{definition} we get
\begin{eqnarray}
F(\phi)=\sum_n T_n = G_T; & I(\phi)=G_T \sin \phi
\label{TJ}
\end{eqnarray}
An alternative, more microscopic approach, would be to
derive boundary condition for $\Lambda(x)$ at the barrier,
which is similar to Eq. \ref{interface}. This work
has been performed by Kuprianov and Lukichev for
superconductivity \cite{Kuprianov}. It has been shown
in \cite{Universality} to lead to the same results as
Eq. \ref{TJ}.

Another important element is a quantum point contact.
\cite{QPC,QPCRus} By virtue of the semiclassical approach we
use we are able to consider only wide QPC, with many propagating
modes, $G \gg G_Q$. For a QPC, all the transmissions are either
zero or unity. Then the functions in use become
\begin{eqnarray}
F(\phi)= \frac{G_{QPC}}{\cos^2(\phi/2)}; I(\phi)= 2 G_{QPC}\tan(\phi/2).
\end{eqnarray}

The important point to be clarified is how to conjugate
these elements. By virtue of the diffusion approach
developed we shall assume that the special
elements are connected with diffusive conductors.
The resistivity of those, however, can be
vanishing in comparison with the resistivity of
the special elements. In this case we can say
the special elements are {\it directly} connected. We shall
assume, however, that there is enough disorder
to prevent the immediate ballistic transmission
from one element to another one. Talking of
microscopic background, the source of such disorder
is not necessarily the scattering at impurities
in the bulk of the material. It can be due to
scattering on boundaries of the conductor, in
the latter case we would have a chaotic ballistic
transmission.

One can think of an imaginable  element which
would provide the crossover between tunnel junction
and ballistic constriction. The simplest model
for it would be to set all transmissions to a certain
value $\Gamma$, $1>\Gamma>0$, as it has been proposed
in \cite{CarloScaling}. It looks like it is a good
model for a semitransparent interface, but one
should be cautious: the study of Bauer \cite{Bauer}
shows that the transmission distribution of an
interface may be very complicated.

\subsection{Rules}
Let us formulate the set of rules which automate
the calculations of transmission distribution
for an arbitrary coherent network of the kind we
describe.
\begin{itemize}
\item{I.} Construct the two terminal network from the elements
which can be diffusive conductors, tunnel junctions
and quantum point contacts. Ascribe the "potential"
$\phi$ to one terminal, zero potential to another one,
"potentials" $\theta_i$ in nodes remain undetermined.
\item{II.} Ascribe the certain $I-\theta$ characteristic
to each element. It relates to its conductivity G. For
  \begin{itemize}
   \item{Diffusive conductor:} $I(\theta)=G \theta$
   \item{Tunnel junction:} $I(\theta)=G \sin \theta$
   \item{QPC:} $I(\theta)=2 G \tan(\theta/2)$
  \end{itemize}
$\theta$ being the difference between "potentials"
of the terminal nodes of the element.
\item{III.} Take into account the current conservation
in the nodes. This supplies the number of equation
sufficient to determine all $\theta_i$.
\item{IV.} Solve the "potential" distribution in the
nodes.
\item{V.} With the aid of this, obtain the total "current" as a function
of $\phi$.
\item{VI.} Compute the transmission distribution
with the aid of Eq. \ref{analytic}.
\end{itemize}

As we will see below, many physical quantities
may be calculated immediately from $I(\phi)$.

\section{Examples and results}
\subsection{Classic results}
Here I will present a small collection of physical results which
can be obtained making use of universal transmission
distribution (\ref{result}). As it is being stressed
throughout the paper, the results are valid for an arbitrary
 diffusive classical conductor. I am tempted to call these
classic rather than classical: the formulas are so elegant
and concise.

Before giving those let us first notice that all the quantities mentioned
in Sec. 2.3 can be expressed in terms of $I(\phi)$ immediately,
without referring to transmission distribution (\ref{result}).
The corresponding relations can be obtained by transformation
of (\ref{definition}) .

The excess noise suppression (\ref{noise}) is given by
\begin{equation}
P/P_{Poisson} = \frac{1}{3} \left( 1- \frac{2 I'''(0)}{I'(0)}\right)
\label{noise_formula}
\end{equation}
Andreev conductivity (\ref{Andreev}) can be expressed as follows
\begin{equation}
G_A =I'(\frac{\pi}{2}),
\label{Andreev_formula}
\end{equation}
and Josephson current (\ref{supercurrent}) appears to be
\begin{equation}
I_{J}(\phi_{sup})= \frac{\Delta \sin \phi_{sup}}{ 2 \sqrt{2} e}
\int_0^{\phi_{sup}} \frac{I(\phi) d \phi}
{\sin(\phi/2) \sqrt{\cos \phi_{sup}-\cos \phi}}.
\end{equation}

For universal distribution, one has to substitute $I(\phi)=G \phi$.
This yields "one-third" suppression \cite{onethird}
\begin{equation}
P/P_{Poisson} = \frac{1}{3}.
\end{equation}
For Andreev conductance, the result can not be simpler:
\begin{equation}
G_A=G.
\end{equation}
This has been proven in
\cite{LarkinOhm} by traditional superconductivity methods.

For superconducting current in SNS system we reproduce
the result of Kulik and Omel'yanchuk, \cite{Kulik}
\begin{equation}
I_J(\phi_{sup})= \frac{\pi \Delta}{e} \cos(\phi_{sup}/2)
 {\rm arctanh} (\sin(\phi_{sup}/2)).
\end{equation}
They indicated an interesting peculiarity: derivative
of the current with respect to the phase diverges
at $\phi_{sup} \rightarrow \pi$. From Eq. \ref{Andreev}
we see the reason for that: nonvanishing transmission
density at $T \rightarrow 1$.

Very interesting results have been recently obtained by Levitov
and co-workers. \cite{chi} They combine the binomial
statistics (\ref{noisestatistics}) and the universal transmission
distribution. They determined the characteristic
function of the number of electrons transmitted through
diffusive conductor over time interval $t$
defined as
\begin{equation}
\chi(\lambda)= \sum_K P_K e^{i \lambda K},
\end{equation}
$P_K$ being the probability to have $K$ electrons transmitted.
They found relatively simple expression
\begin{equation}
\chi(\lambda)=\exp \left( \frac{It}{e}
{\rm arcsinh}^2 \sqrt{e^{i\lambda}-1}\right)
\end{equation}
$I$ being the average current passing the circuit.

Their expression can be generalized for
an arbitrary circuit \cite{tobepub}:
\begin{equation}
\ln \chi(\lambda)= \frac{I t}{2 e}
\int_0^{2 {\rm arcsinh}^2 \sqrt{e^{i\lambda}-1}}
I(\phi) d \phi.
\end{equation}

\subsection{Tunnel junction and diffusive conductor}
Let us apply the circuit theory to the simplest example.
Let us have a diffusive conductor,
tunnel junction and another diffusive conductor in series.
Although not the most general, this
setup is relevant for quasi-one-dimensional geometry,
sandwich type geometry and for any situation when the equipotential
surface of the voltage distribution coincides with the tunnel
interface.
We shall match the
total currents through the leftmost diffusive conductor,
the tunnel junction and the rightmost conductor. These
currents are related to drops of $\theta$ at every element
by means of the circuit theory rules:
\[ I =
G_{Nleft}\theta_1=G_T\sin(\theta_2-\theta_1)=G_{Nright}(\phi-\theta_2).\]
$\theta_{1,2}$ being "potentials" respectively on the left and on the
right side of the tunnel junction.
This is enough to express the current in terms of $\phi$,
\begin{equation}
I(\phi) = G_N {X(\phi)};
X(\phi)+\frac{G_T}{G_N} \sin(X(\phi)-\phi))=0.
\label{iphi_tj}
\end{equation}
Here $G_N$ stands for the total conductance of two diffusive parts,
$G_N^{-1}= G_{Nleft}^{-1}+G_{Nleft}^{-1}$.
To extract the distribution, let us make use of
(\ref{definition}). The answer comes in an implicit form
which suffice to analyze it,($\alpha \equiv G_T/G_N$)
\begin{eqnarray}
\rho(T)=f\rho_0(T); \\
T={\rm cosh}^{-2}\mu\\
\mu= \frac{1}{2} \left( {\rm arccosh}(\frac{\pi f}{\alpha\sin(\pi f)})-
\alpha \sqrt{(\frac{\pi f}{\alpha\sin(\pi f)})^2-1} \cos(\pi f)\right).
\end{eqnarray}

The distribution is always suppressed in comparison with the universal value.
The more resistive is the barrier, the larger is the suppression.
If $G_T>G_N$, the distribution remains finite at $T\rightarrow1$,
this indicates that a certain fraction of the channels have almost
absolute transmission. This fraction declines to zero at $G_T=G_N$.
At $G_T>G_N$ the maximal transmission available can not exceed a certain value
$T_{max}$, $T_{max}=1,G_N=G_T$; $T_{max}=4 G_T^2/G_N \ll 1$, $G_T \ll G_N$.

These features reveal an important new physics which is absent
if tunneling occurs between two clean metals. In the latter case,
transmissions through the barrier are always restricted by some
maximal value depending on the structure of the barrier.
Intuitively, one could expect that the same value restricts
the maximal transmission through the compound system of the tunnel
barrier and diffusive conductor. It is not the case. The average
transmission is suppressed by an addition of extra scatterers
but some channels become even more transparent having transmissions
of the order of unity. This can be comprehended as a result
of constructive quantum interference between different trajectories
traversing the barrier. If the metals are clean, a typical
trajectory reaches the barrier only once either reflecting or getting through.
If the metals are disordered, the typical trajectory gets back to
the barrier if it was reflected initially. This enhances the possibilities
for interference.

The channels with transmissions close to unity have been associated
with delocalized states.\cite{openchannels} If the system under consideration
were uniform, the disappearance of the delocalized at the point $G_N=G_T$ would
have
meant the true localization transition. It is clear that something
drastic happens at this point, but one should be cautious drawing
direct conclusions. We would stress that the mere introduction
of the transmission distribution by (\ref{Landauer}) is an attempt
to describe the almost classical system in quantum mechanical terms.
This can make the results difficult to interpret.
Indeed,
neither resistance
\[R=R_T+R_N\]
nor excess noise power
 \[P=\frac{P_{classical}}{3} (1+2(\frac{R_T}{R})^3)\]
in the system exhibit critical behavior around the
transition point within the framework of the semiclassical approach used.
Such behavior could  emerge from quantum localization corrections.
The Josepson current appears to be sensitive to
the transition. Indeed, as we mentioned before,
the finite transmission density results in
divergent current derivative at $\phi_sup=\pi$.
On the localized side of the transition there
is no such peculiarity.

Eqs. \ref{Andreev_formula} and \ref{iphi_tj} give the Andreev
conductance of the system
\[G_{A}=G_N/(1+{\cot}(X)/X),\cos(X)/X=G_N/G_T ;\]
this coincides with the result of \cite{Klapwijk} in
the relevant limit.
\subsection{Two tunnel junctions}
We consider next  two tunnel junctions  in series.
As it was explained above we assume diffusive motion
of electron between the junctions, albeit the resistance
of the disordered region is supposed to be much smaller
than the resistance of the junctions.

We find the "current" by equating the "currents" through
each junctions, $\theta$ being the "potential" of the middle,
$G_{1,2}$ being junction conductances,
\begin{eqnarray}
I(\phi)=G_1 \sin(\theta)= G_2 \sin(\phi-\theta);\\ \nonumber
I(\phi)=\frac{G_1 G_2 \sin\phi}{\sqrt{G_1^2+G_2^2+2 G_1 G_2 \cos\phi}}
\label{iphi_2tj}
\end{eqnarray}

Performing analytical continuation (\ref{analytic}),
we readily find the transmission distribution:
\begin{eqnarray}
\rho(T)=\frac{G}{2 \pi G_Q} \frac{1}{\sqrt{T (T_c-T)}}, &
T <T_c, T_c \equiv \frac{4 G_1 G_2}{(G_1+G_2)^2}
\end{eqnarray}
First, this transmission distribution differs
drastically from that one for a single tunnel
barrier being spread at $T \simeq 1$. This is due to resonant transmission
via electron states localized between junctions.
Second, it always has a gap at $T$ close to unity
if the junction conductances do not match.
We shall conclude that the equal probabilities
to escape to each side provide a necessary condition
for the resonant state to transmit with unitary
probability.

{}From Eqs. \ref{noise_formula}, \ref{Andreev_formula}
and Eq. \ref{iphi_2tj} we obtain noise suppression
and Andreev conductance
\begin{eqnarray}
P/P_{Poisson}=1-\frac{G_1 G_2}{(G_1+G_2)^2}, \ \ \
G_A= \frac{G_1^2 G_2^2}{(G_1^2+G_2^2)^{3/2}}.
\end{eqnarray}

\subsection{Two point contacts}
 Now we consider transmission through two point contacts.
 If there is a possibility of a direct ballistic
transmission through both of them,
 the total conductance is determined by a
maximal number of transmitted channels, $G=min\{G_1,G_2\}$.
Otherwise  Ohm's law $R=R_1+R_2$ is restored and
we are able to apply the circuit theory.

Again we find $I(\phi)$ by equating "currents":
\begin{eqnarray}
I(\phi)/2=G_1 \tan(\theta/2)= G_2 \tan((\phi-\theta)/2);\\ \nonumber
I(\phi)=(G_1 +G_2)\cot(\phi/2)\{\sqrt{1+
\frac{4 G_1 G_2}{(G_1+G_2)^2}\tan^2(\phi/2)}-1 \}.
\end{eqnarray}
This gives the following transmission distribution:
\begin{eqnarray}
\rho(T)= \frac{2(G_1+G_2)}{\pi G_Q T} \sqrt{\frac{T-T_c}{1-T}} & {\rm at} \
T>T_c; &
T_c=(\frac{G_1-G_2}{G_1+G_2})^2,
\end{eqnarray}
$\rho(T)=0$ otherwise.
It is almost exactly reversed if compared
with two tunnel junction case. Again, the distribution is
spread to $T \simeq 1$. But it has  a gap at small
transmissions. The gap is closed if the two QPC conductances
match. If one of the conductances is much bigger than another one,
the distribution shrinks  to $T=1$ approaching a single
QPC distribution.

Noise suppression and Andreev conductance for this case
read
\begin{eqnarray}
P/P_{Poisson}=\frac{G_1 G_2}{(G_1+G_2)^2}, \ \
G_A= (G_1+G_2)(1-\frac{G_1+ G_2}
{\sqrt{G_1^2+G_2^2+6 G_1 G_2}}).
\end{eqnarray}

Two QPC in series may serve as a simplest microscopic
model of so-called chaotic cavity. In such a system,
the scattering at the boundaries only is important. At arbitrary
shape of the boundaries, a particle entering the cavity
moves chaotically and experiences many collisions before
leaving it.

It has been conjectured in
\cite{CavityMello,CavityPichard}
that the scattering matrix of the system is a member
of Dyson circular ensemble. This is a typical
judgment of random matrix theorists about real things:
 it is a bolt from the blue and does not look provable.
Let us see if we can check it. For the Dyson
circular ensemble, in the semiclassical limit
of large number of channels, the transmission
distribution reads \cite{CavityPichard}
\begin{equation}
\rho(T)=\frac{G}{\pi G_Q} \frac{1}{\sqrt{T(1-T)}}
\end{equation}
in our notations.
It is precisely what we got if $G_1=G_2$! It is a strong
argument in support of the conjecture.
However, for more general situation $G_1 \ne G_2$ we have
qualitatively different transmission distribution. The random matrix
theory obviously skips some interesting physics being
insensitive to such "details".

I wish to thank C. W. J. Beenakker, L. S. Levitov and D. E.
Khmelnitskii. I greatly profited
from each discussion with those.

\end{document}